\documentclass[aps,prl,showpacs,twocolumn,reprint,amsmath,amssymb]{revtex4-1}
\usepackage{amsmath}
\usepackage{graphicx}
\begin{document}
\title{Revealing the substrate origin of the linear dispersion of silicene/Ag(111)}
\author {M. X. Chen}
\affiliation{Department of Physics, University of Wisconsin, Milwaukee, Wisconsin 53211, USA}
\author {M. Weinert}
\affiliation{Department of Physics, University of Wisconsin, Milwaukee, Wisconsin 53211, USA}

\date{\today}

\begin{abstract}
The band structure of the recently synthesized (3$\times$3) silicene
monolayer on (4$\times$4) Ag(111) is investigated using density
functional theory. A $k$-projection technique that includes the
$k_\bot$-dependence of the surface bands is used to separate the
contributions arising from the silicene and the substrate, allowing
a consistent comparison between the calculations and
the angle-resolved photoemission experiments.  Our calculations
not only reproduce the observed gap and linear dispersion across the K point of
(1$\times$1) silicene, but also demonstrate that these originate from the
$k_\bot$-dependence of
Ag(111) substrate states (modified by interactions with the silicene)
and \textit{not} from a Dirac state.
\end{abstract}


\maketitle
The recent theoretical prediction of silicene, the silicon counterpart
of graphene, has generated intense interest in this new two-dimensional
system \cite{cahangirov_two-_2009}.  Unlike graphene, silicene is
predicted to have a low-buckled honeycomb structure, in which one Si
atom is buckled out of the plane, but the linear band dispersion at the
K point -- the Dirac states -- is preserved by the crystal symmetry.
While graphene can be mechanically exfoliated from graphite, silicene
has to be grown on a substrate.\cite{PhysRevLett.108.245501,lalmi_epitaxial_2010,0953-8984-24-17-172001} 
Recently, silicene monolayers were
epitaxially grown on Ag(111)
\cite{lalmi_epitaxial_2010,0953-8984-24-17-172001,vogt_silicene:_2012,chen_evidence_2012,
feng_evidence_2012,fleurence_experimental_2012,chen_spontaneous_2013,meng_buckled_2013} and 
various structures were reported by scanning tunneling microscopy (STM)
experiment, such as (3$\times$3) \cite{vogt_silicene:_2012} and
($\sqrt{3}$$\times$$\sqrt{3}$) \cite{chen_evidence_2012} reconstructions
with respect to ideal silicene.  The existence of Dirac states in
silicene/Ag(111), however, remains hotly debated.  Angle-resolved
photoemission spectroscopy (ARPES) experiments for (3$\times$3) silicene
on (4$\times$4) Ag(111) show a linear dispersion and a gap forming
around the K point of (1$\times$1) silicene \cite{vogt_silicene:_2012}.
Linear dispersion above the Fermi level ($E_F$) was also deduced
from STM experiments \cite{chen_evidence_2012,chen_spontaneous_2013} for the
($\sqrt{3}$$\times$$\sqrt{3}$) reconstruction of silicene/Ag(111).  In
contrast, the Landau levels for (3$\times$3)-silicene on (4$\times$4)
Ag(111) [henceforth referred to as silicene/Ag(111)] in  a magnetic
field are distinct from those of graphite, suggesting the absence of a
Dirac point in the system \cite{lin_substrate-induced_2013}.

There have been a number of density functional theory (DFT) calculations
to clarify whether the Dirac point exists in silicene/Ag(111) and
explain the nature of the linear band observed by APRES experiment
\cite{guo_absence_2013-1,guo_absence_2013,wang_absence_2013,gori_origin_2013,
cahangirov_electronic_2013,quhe_does_2014,mahatha_silicene_2014}.
The linear dispersion was not seen in these DFT calculations. Instead,
Refs.~\onlinecite{wang_absence_2013} and \onlinecite{gori_origin_2013}
show that a Ag band has similar dispersion to the linear band
observed by ARPES, and Ref.~\onlinecite{cahangirov_electronic_2013}
claims that the linear dispersion results from the hybridization between
silicene and the substrate since such a state disappears in the pure substrate. 
Further ARPES
experiments\cite{avila_presence_2013,tsoutsou_evidence_2013,mahatha_silicene_2014}
observe similar spectral features, but have attributed the linear
dispersion to
gapped silicene bands\cite{avila_presence_2013}, 
hybridized metallic surface states\cite{tsoutsou_evidence_2013}, 
or silicene-induced Ag free-electron states.\cite{mahatha_silicene_2014}
Despite the considerable research devoted to this system, attempts to establish a link between
the DFT calculations and the experimental observations are at present
incomplete since the calculations have not been able to provide a
consistent explanation of the experimental results for both the clean
substrate and the silicene/Ag(111) system. 

In this paper, we use a $k$-projection unfolding
scheme \cite{kprojCu3Au,bufferlayer} to reconcile the experimental
observations and the calculations for both the pure Ag(111) substrate
and silicene/Ag(111).
We demonstrate that the linear
dispersion is not due to Dirac states of the silicene, but rather due to
substrate states that are modified by the interaction with the silicene.
The experimental observations are demonstrated to be consistent with the
$k_\bot$-dependence of the bands probed by the photoemission experiments.

In our calculations, the Ag(111) surface substrate is modeled by a
ten-atomic-layer slab, which is separated from its periodic images by
$\sim$20 \AA{} vacuum regions.  Silicene monolayers are
symmetrically placed on both sides of the substrate slab to to avoid
dipole interactions between slabs.  The electronic and structural
properties are calculated using the local density approximation and
VASP \cite{kresse_efficiency_1996,kresse_efficient_1996} with the
projector augmented wave potentials to represent the ion cores.  The
surface Brillouin zone (BZ) is sampled by $k$-point meshes that are
equivalent to the 6$\times$6 $\Gamma$-centered Monkhorst-Pack mesh for a (4$\times$4)
cell.  Atoms in the middle eight layers of the Ag substrate are frozen
at the bulk geometry, while all other atoms are fully relaxed until the
residual forces are less than 0.001 eV/\AA{}. The influence of
van der Waals (vdW) dispersion forces between the adsorbate and the substrate 
were examined using dispersion-corrected DFT-D3
calculations\cite{Grimme}:
the additional forces (and subsequent changes in atomic positions) were
found to be within the force convergence criteria, reflecting the
relative importance of the direct Si-Ag bonding.

The unfolded bands are obtained using the $k$-projection method
\cite{kprojCu3Au,bufferlayer} in which the projection of a function $\psi$ that
transforms as the irreducible representation of the translation group
labeled by $\mathbf{k}$ is given by
\[
 \psi_\mathbf{k} = \hat{P}_\mathbf{k}\, \psi = \frac{1}{h} \,
\sum_\mathbf{t} \chi^*_\mathbf{k}(\mathbf{t}) \hat{T}_\mathbf{t} \psi ,
\]
where $\hat{T}_\mathbf{t}$ is the translational operator corresponding
to the translation $\mathbf{t}$ with  character
$\chi_\mathbf{k}(\mathbf{t}) =  e^{i\mathbf{k}\cdot\mathbf{t}}$, and
where the set $\lbrace\mathbf{t}\rbrace$ of order $h$ corresponds to the
translations associated with a cell defined by direct and reciprocal
lattice vectors $\mathbf{a}_i$ and $\mathbf{b}_j$, respectively. In
practice, $\psi$ is most often a wave function calculated at
$\mathbf{k}_s$ of a supercell (defined by lattice vectors $\mathbf{A}_i$
and $\mathbf{B}_j$), and $\psi_\mathbf{k}$ is the projection on to the
primitive cell.  (The projection may be on to an even smaller cell:
for example, in the case of Cu$_3$Au \cite{kprojCu3Au} the unit
cell is simple cubic, but an fcc ``primitive'' cell is appropriate for 
describing the Cu bands.)

For plane-wave-based representations of the wave functions with commensurate primitive and
super-cells, this
procedure is particularly simple, reducing to the problem of determining
which $\mathbf{k}_p$ each plane wave $e^{i\mathbf{G}\cdot\mathbf{r}}$
belongs to, i.e., for integers $M_i$ and $m_j$, determining the fractional part $\kappa_j$ that
defines $\mathbf{k}_p$ of the primitive cell relative to $\mathbf{k}_s$
of the supercell: 
\begin{eqnarray*}
 \mathbf{G} &=& \sum_i M_i \mathbf{B}_i  = \sum_j (m_j + \kappa_j ) \mathbf{b}_j \\
            &=& \sum_j \left( \sum_i M_i (
\mathbf{B}_i\cdot\mathbf{a}_j ) \right)  \mathbf{b}_j ,
\end{eqnarray*}
with $\mathbf{a}_i\cdot\mathbf{b}_j=\delta_{ij}$.  For an ideal
supercell, this decomposition is exact since it is a simple consequence
of translational symmetry and recovers the primitive band structure with
$\langle\psi_{\mathbf{k}_p}|\psi_{\mathbf{k}_p}\rangle=1$ or 0, while for
defect systems, the norm will be between zero and one.  
Once a wave function has been $k$-projected, its weight
$|\psi_\mathbf{k}(\mathbf{r})|^2$ in different spatial regions (or over
all space)  can be obtained by straightforward integration, 
 which is particularly useful for interface structures.
As shown below for silicene/Ag(111), 
a layer projection along $z$ separates the $k$-projected band structure
for the silicene overlayer and the Ag substrate.

\begin{figure}
  \includegraphics[width=0.48\textwidth]{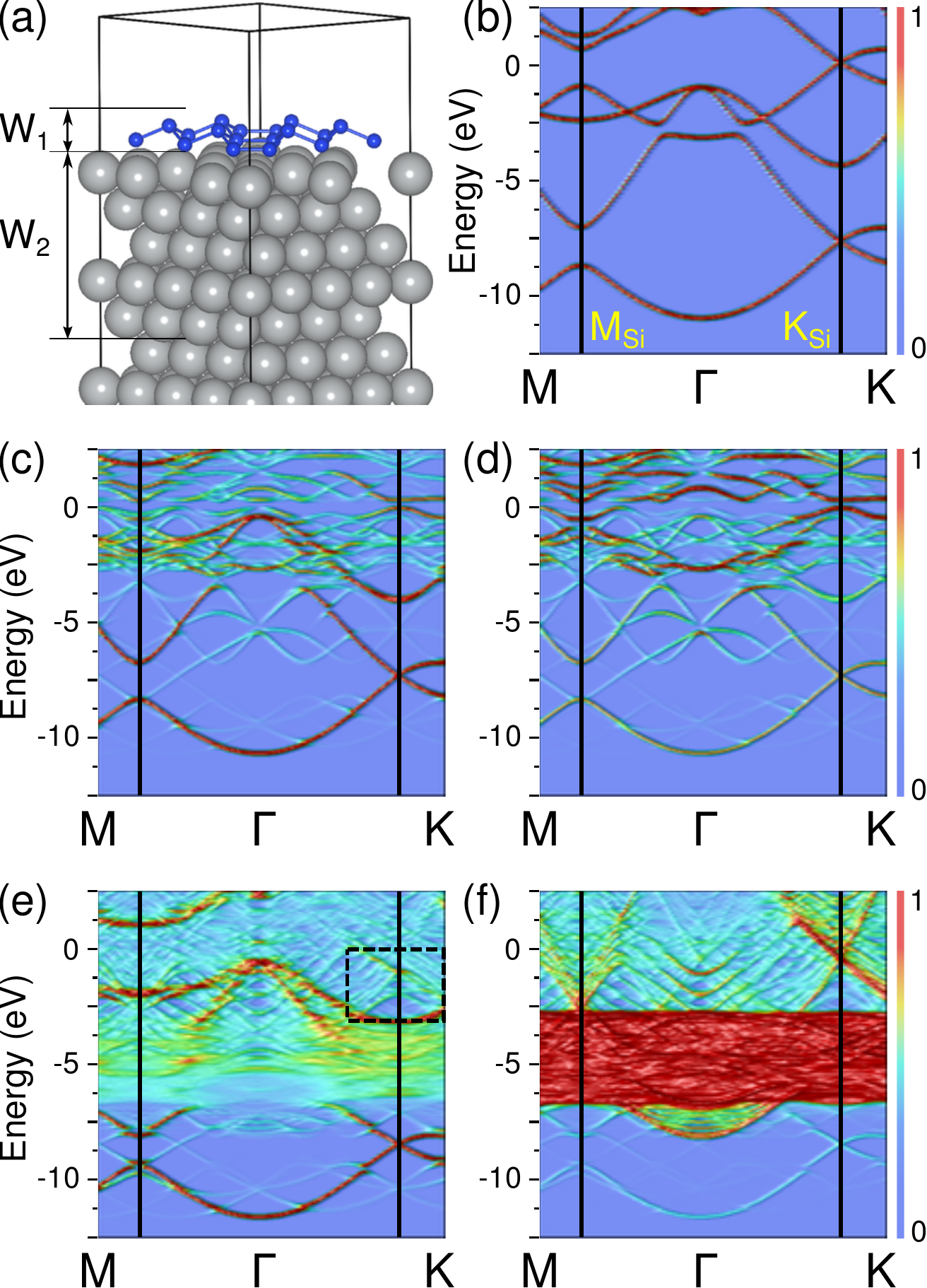}
  \caption{ 
(a) Perspective view of the relaxed structure of
(3$\times$3) silicene on (4$\times$4) Ag(111).  W$_1$ (centered on the
silicene) and W$_2$ (substrate) are the spatial regions used for integrating wave functions in $k$-projection. 
$k$-projected bands for
(b) ideal free-standing silicene;
for free-standing distorted silicene, weighted by the contributions (c) in
the silicene layer (W$_1$) and (d)
in the vacuum region above the layer;
and for silicene/Ag(111), weighted by the wave function contributions in (e)
W$_1$ and (f) W$_2$.
K and M (K$_\text{Si}$ and M$_\text{Si}$) correspond to the high
symmetry points of the (1$\times$1) Ag (silicene) surface Brillouin
zone, and $E_F=0$.  The black dashed box in (e) indicates the
experimental window probed by ARPES \cite{vogt_silicene:_2012}.
The color bars indicate the layer- and
$k$-projected weights of the bands relative (between 0 and 1) to the
maximum in the plot; the same color scheme is used in subsequent figures
also.}
 \label{fig1}
\end{figure}

Our relaxed silicene/Ag(111) structure, Fig~\ref{fig1}(a), is distorted
compared to ideal free-standing silicene: six atoms that reside above Ag
atoms are shifted up, resulting in a new structure that has a mirror
symmetry about the (110) plane.  Moreover, the buckling ($\sim$0.9\AA)
is significantly enhanced in silicene/Ag(111), almost twice that of the
ideal silicene, implying strong perturbations of the silicene bands due
to interactions with the substrate.  The simulated STM image for this
structure is in good agreement with experiments and previous DFT
simulations \cite{vogt_silicene:_2012,lin_substrate-induced_2013,guo_absence_2013-1,guo_absence_2013,cahangirov_electronic_2013}.
Because of this reconstruction, the degeneracy and linear dispersion at
K$_\text{Si}$ seen for the ideal (1$\times$1) silicene,
Fig.~\ref{fig1}(b) is no longer required by symmetry.

Although the distorted silicene no longer has the (1$\times$1) silicene
periodicity, the bands can still be unfolded into the silicene BZ by
projecting the supercell wave functions onto the corresponding $k$ of
the (1$\times$1) silicene cell. 
Figures\ref{fig1}(c,d) depict the $k$-projected band structure of the distorted
silicene without the substrate, where the $k$-projected bands are
weighted by a layer-integration over the wave functions in the silicene
layer (Fig.~\ref{fig1}(c)) and the vacuum region (Fig.~\ref{fig1}(d)) to
facilitate investigation of silicene-substrate interaction to be discussed later.  
While the $\sigma$ bands of the ideal silicene are fairly
well preserved in the distorted structure, the linear $\pi$ and $\pi*$
bands around the K$_\text{Si}$ are strongly perturbed, leading to a gap
opening at the Fermi level because of the symmetry breaking associated
with the (3$\times$3) structural reconstruction, consistent with previous calculations\cite{pflugradt_2014}.  
Because the silicene $\pi$ orbitals extend farther into the vacuum than the $\sigma$ ones,
they will have more overlap, and hence interaction, with the Ag
substrate states.

For the interacting (distorted) silicene-Ag(111) system, the bands
$k$-projected onto (1$\times$1) silicene, Fig.~\ref{fig1}(e), show that
the $\sigma$ bands of silicene are similar, although shifted
down in energy by $\sim$1--2 eV. The $\pi$ states, however, are strongly
modified, with the result that the (gapped) $\pi$ states in the isolated
(distorted) silicene around K$_\text{Si}$ are indistinct in the $E$
vs.\ $k$ window corresponding to the ARPES experiments indicated by the
black box.

\begin{figure}
  \includegraphics[width=0.48\textwidth]{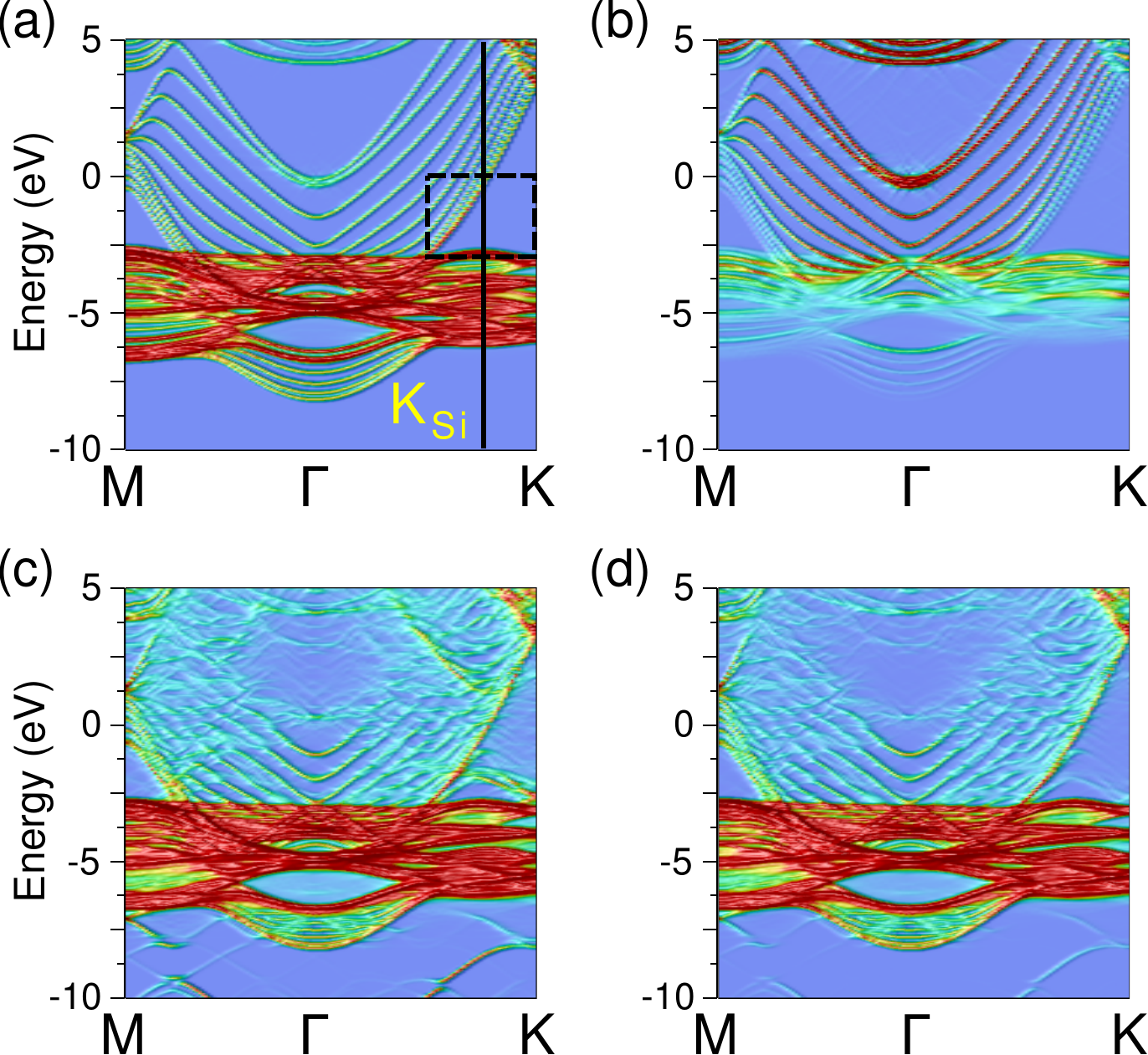}
  \caption{
Bands $k$-projected to the (1$\times$1) Ag(111) for:  the Ag(111)
surface in (a) the substrate (W$_2$) and (b) vacuum (W$_1$); and
for silicene/Ag(111) in (c)
both the silicene and the substrate, W$_1$ + W$_2$, and (d) the
substrate only, W$_2$.  The dashed black box in (a) indicates the
experimental window of ARPES \cite{vogt_silicene:_2012}.  Solid black
lines mark the K$_\text{Si}$ point, and $E_F=0$. 
  }
 \label{fig2}
\end{figure}

The bands projected on to (1$\times$1) silicene but weighted by their
contribution in the
substrate, Fig.~\ref{fig1}(f), are dominated by the Ag $d$ bands around
$-$5 eV and the free-electron $sp$ bands. (Because the projection is to
the silicene cell, rather than (1$\times$1) Ag, the $d$ states do not
project to specific $k$ values, but rather almost uniformly in $k$.
Projecting to the Ag cell as in Fig.~\ref{fig2} recovers the momentum
resolution; the $sp$ states, being free-electron-like,
are not sensitive to the $k$-projection used.) The two nearly linear
Ag-derived bands intersecting at K$_\text{Si}$ near the Fermi level
reflect the folding of the bands due to the (3$\times$3) periodicity
associated with the silicene overlayer, not a symmetry-dictated Dirac
cone.  Similar unfoldings were done in Refs.~\onlinecite{wang_absence_2013}
and \onlinecite{cahangirov_electronic_2013} for silicene/Ag(111), but
the remaining band foldings hid the substrate nature of bands.  

Figure~\ref{fig2} shows $k$-projected bands of Ag(111) without and with
silicene, for the $k$-projection done with respect to (1$\times$1)
Ag(111). Since now the projection is to the ``correct'' reference, the
dispersion of the Ag $d$ bands is well-defined. Comparison of
Figs.~\ref{fig2}(a) and (c) shows essentially no changes in the Ag $d$
bands when the silicene is adsorbed. Because of the finite
number of layers in the Ag substrate, the $sp$ bulk band is represented
by a set of 10 bands in Fig.~\ref{fig2} which have
spectral weight within the  $E$ vs.\ $k$ window probed
experimentally \cite{vogt_silicene:_2012}, and thus should be accessible
to ARPES measurement. In particular, the band edge state crosses
$K_\text{Si}$ with almost linear dispersion. By comparing the
spatial distributions of the states in Figs.~\ref{fig2}(a,b),
the higher energy $sp$ states have greater surface weight, suggesting
that these states will have the greatest overlap with the silicene.

For the silicene/Ag(111) case, Figs.~\ref{fig2}(c,d), a comparison of
the dispersion and intensity of bands with those of the
Ag(111) surface indicates that the observed linear band is predominately
derived from substrate. Comparing Fig~\ref{fig2}(d) with
Fig~\ref{fig1}(d), the crossing at the K$_\text{Si}$ point disappears when
the bands are completely unfolded, confirming the absence of the Dirac
cone in silicene/Ag(111).

We now discuss the relation between our DFT calculations and the
experimental observations.  Because the wave functions of
silicene/Ag(111) are dominated by silicene in W$_1$ and by the substrate
in W$_2$, respectively, a reasonable approximation to the experiment
observation is to use the $k$-projected bands appropriate to the
underlying translational symmetry of the different spatial regions,
i.e.,  for silicene, W$_1$ [Fig~\ref{fig1}(e)], and for the substrate,
W$_2$ [Fig~\ref{fig2}(d)].  The calculations thus show that
the linear band in
Fig~\ref{fig2}(d) due to the substrate --- and not a silicene band ---
should be identified with the one observed by ARPES. 

A consistent interpretation to the experimental observations requires
explanations for the origin of the linear band across the K$_\text{Si}$; for
the parabolic bands near the M and $\Gamma$ points; for the gap opening
at the K$_\text{Si}$, M, and $\Gamma$ points for silicene/Ag(111); and the
absence of bands in the experimental range for the pure
substrate \cite{vogt_silicene:_2012,avila_presence_2013}. Previously,
Ref.~\onlinecite{cahangirov_electronic_2013} attributed the linear band
observed by ARPES to  a new hybridized state due to the strong
silicene-substrate interaction because it disappears in pure Ag(111),
while our calculations suggest that it is a (modified) substrate state,
either a surface resonance near the bottom of the $sp$ band or a true
surface state split off from the bottom of the band. (Because of the
finite number of layers used to represent the substrate, determining
whether this lowest band is at or just below the band edge is difficult to
determine; the increased localization in the surface region compared to
the clean substrate is consistent with either case.)
Moreover, whether a state is observed or not in ARPES is not simply a
matter of whether or not it is hybridized.
Refs.~\onlinecite{wang_absence_2013} and \onlinecite{gori_origin_2013}
show that the linear band observed by ARPES looks similar to one of Ag
bands in the projected effective band structure, 
but the relationship between the experiments and calculations was not explored.

Our results for the pure substrate indicate that there are some Ag bands
in the experimental range [Fig.~\ref{fig2}(a)] which were, however, not
observed by APRES.  Because these bands are bulk bands, they have a
strong $k_\bot$-dependence due to the large band widths (cf.,
Fig.~\ref{fig2}(a)).  By choosing particular photon energies, the ARPES
experiments \cite{vogt_silicene:_2012} could choose a window with no bulk
Ag states visible. To make quantitative connection with the experiments,
we thus need to account for the experimental conditions.

We address these issues by including in the $k$-projection
the momentum perpendicular to the surface, $k_\bot$, probed by the
photoelectron, i.e.,
\begin{equation}
 k_{\bot} = \sqrt{\frac{2m_e}{\hbar^2}(h\nu-\phi-|E_B|)-\mathbf{k}_{\shortparallel}^2},
\end{equation}
where $h\nu$ is the photon energy, $\phi$ is the work function of the
system, $E_B$ is the binding energy of electrons, and
$\mathbf{k}_{\shortparallel}$ is the component parallel to the surface
of the electron crystal momentum (with both $k_\bot$ and
$\mathbf{k}_\shortparallel$ determined up to a reciprocal lattice vector).
Our calculated work functions for the pure substrate and
silicene/Ag(111) are 4.78 and 4.67 eV, respectively.  The range of
$|E_B|$ for the ARPES experiments in
Ref.~\onlinecite{vogt_silicene:_2012} is 3 eV and 1.5 eV for those in
Ref.~\onlinecite{avila_presence_2013}.  For
$\mathbf{k}_{\shortparallel}$ = K$_\text{Si}$ and a photon energy of 126 eV
as in the experiments, $k_{\bot}$ is in the range of 0.7--1.1
\AA{}$^{-1}$ for the pure substrate, while
for silicene/Ag(111) they are 2.03--2.27 \AA{}$^{-1}$ along $\Gamma$-K and
0.37--0.7 \AA{}$^{-1}$ for $\Gamma$-M-$\Gamma$ 
(taking reciprocal lattice vectors into account for the estimates).

\begin{figure}
  \includegraphics[width=0.48\textwidth]{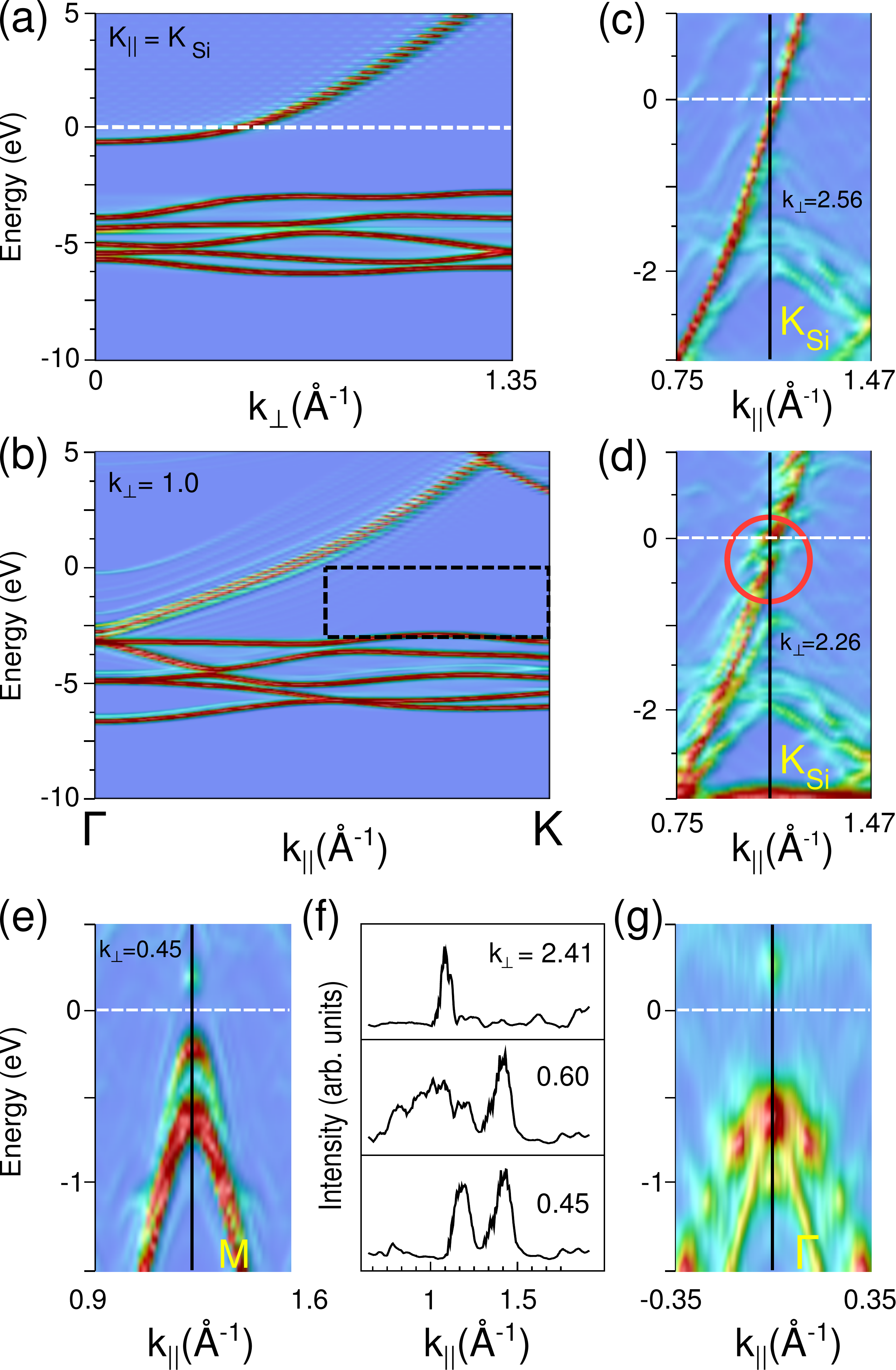}
  \caption{(a) E($k_{\bot}$) for $k_{\shortparallel}$ = K$_\text{Si}$ and 
(b) E($k_{\shortparallel}$) along $\Gamma$-K with $k_{\bot}$ =1.0
\AA{}$^{-1}$ for a 40-layer Ag(111) film.
The black dashed box is the experimental window~\cite{vogt_silicene:_2012}.
(c) E($k_{\shortparallel}$) around K$_\text{Si}$ for silicene/Ag(111) along $\Gamma$-K for
$k_{\bot}$ = 2.56 \AA{}$^{-1}$  and (d)  2.26 \AA{}$^{-1}$.
(e) E($k_{\shortparallel}$) around M (along the $\Gamma$-M-$\Gamma$
direction) for silicene/Ag(111) with $k_{\bot}$ =0.45 \AA{}$^{-1}$.  
(f) Simulated momentum distribution curves along the $\Gamma-M$
direction for $E_B$ = 1.0 eV and for $k_{\bot}$=2.41, 0.60, 0.45
\AA{}$^{-1}$, corresponding approximately to 85, 105, and 126 eV photon
energies, respectively. (g) $k$-projected bands around $\Gamma$ for silicene in
silicene/Ag(111) along K-$\Gamma$-K, weighted by contributions on Si
atoms only.
The $k$-projections for (a)--(f) were done with respect to the bulk Ag unit cell, $E_F=0$.}
 \label{fig3}
\end{figure}

Figure \ref{fig3}(a) shows E($k_{\bot}$) for the Ag(111) surface in the first BZ
corresponding to fcc Ag at $\mathbf{k}_{\shortparallel}$ = K$_\text{Si}$ \cite{primitive_Ag111}.
Not surprisingly, these are essentially the bulk bands for this
$\mathbf{k}_\shortparallel$, with the effect of the surface showing up in the
apparent increased widths of some bands and the faint weights corresponding to
the energies of the band extrema. The band of particular interest is the
uppermost one
that starts out below $E_F$ for small $k_\bot$ and then disperses
above for larger $k_\bot$. Thus this band will be seen in ARPES only for
photons corresponding to small $k_\bot$. The $k$-projected surface bands
for $k_\bot$=1 \AA{}$^{-1}$, a value corresponding roughly to the experiment,
are shown in Fig.~\ref{fig3}(b).  In agreement with the
experiment \cite{vogt_silicene:_2012}, there are no Ag states seen in the
selected energy-momentum window.

Similarly, calculated E($k_{\shortparallel}$) around K$_\text{Si}$ for silicene/Ag(111)
for different values of $k_{\bot}$ are shown in Figs.~\ref{fig3}(c,d).
For $k_{\bot}$ = 2.56 \AA$^{-1}$ outside
the estimated experimental range of $k_\bot$, there is a band crossing
K$_\text{Si}$ that remains continuous throughout the whole $k_{\shortparallel}$ window
[Fig.~\ref{fig3}(c)].  For $k_{\bot}$ = 2.26 \AA{}$^{-1}$, which is in
the estimated range of the experiment, the band shifts up in energy, and
is more diffuse but also more nearly linear.  Moreover, a gap of about
0.3 eV is opened just below $E_F$ at K$_\text{Si}$, in
surprisingly good agreement with the ARPES experiment \cite{vogt_silicene:_2012}.
The $k_\bot$-dependence is also consistent with more recent ARPES
experiments \cite{tsoutsou_evidence_2013,mahatha_silicene_2014}
(c.f., Fig.~3 in Ref.~\onlinecite{tsoutsou_evidence_2013}, and the lack
of an observed gap at K$_\text{Si}$ for the photon energies used in
Ref.~\onlinecite{mahatha_silicene_2014}), both for the
clean Ag substrate and for the silicene/Ag system.
This $k_\bot$-dependent gap opening has its origin in the changed boundary conditions that the Ag $sp$ states
see due to, for example, the presence of (and hybridization with) the silicene layer and changes
in the work function, that will cause modifications to these wave functions in the near surface region.
Moreover, the Fermi velocity derived from the calculated band structure
is $\sim$1.3~$\times$~10$^6$ ms$^{-1}$, 
in good agreement with the experimental value\cite{vogt_silicene:_2012}, 
further indicating the substrate origin of the observed linear dispersion. 

Good agreement between our calculations and
experiment\cite{avila_presence_2013,mahatha_silicene_2014} is also obtained
for the bands about the M point, Fig.~\ref{fig3}(e): the band is
parabolic and separated from the Fermi level by a gap, comparable to the
experimental value.  (The band splitting at M in our
calculation is due to the limited number (10) of Ag layers in
the structural model.) Simulations of momentum distribution curves (MDC)
along $\Gamma$-M-$\Gamma$ were carried out for the $k$-projected bands
of silicene/Ag(111).  Our results for the MDCs around M,
Fig.~\ref{fig3}(f), show the same trend with $k_\bot$ (photon energy) as
the ARPES experiments in Ref.~\onlinecite{avila_presence_2013}; that
the correspondence is not perfect between the experimental photon
energies \cite{avila_presence_2013} and our estimated values of
$k_\bot$, is due in part to differences in work functions.

Bands near $\Gamma$ are complicated by the remaining band folding of the
Ag-derived bands noted in Fig.~\ref{fig1}(f). To eliminate these Ag
contributions, the $k$-projected bands are weighted by the partial
density of states on the silicon atoms only.  Bands along K-$\Gamma$-K
are shown in Fig.~\ref{fig3}(g), which shows two silicene bands at about
$-$0.5 eV at $\Gamma$, with the substrate bands (cf.,
Fig.~\ref{fig2}(c)) are further below starting at about $-$1 eV. 
In the experiments, however,
 only one silicene band with asymmetric intensity with
respect to $k$ was seen [Fig.~4(b) of
Ref.~\onlinecite{avila_presence_2013}], which we tentatively attribute
to polarization and
matrix elements effects similar to those demonstrated for
Cu(111) \cite{mulazzi_understanding_2009}.  While a more complete
simulation of ARPES is beyond the scope of the present work, our results
nevertheless indicate that all the bands about $\Gamma$ in the energy
window of the ARPES experiment are parabolic [Figs.~\ref{fig2}(c) and
~\ref{fig3}(g)] and are not related to Dirac states at K$_\text{Si}$
folded back to $\Gamma$.

In summary, based on first-principles calculations of the $k$-projected
bands we have elucidated the origin of the linear dispersion
observed by ARPES for silicene/Ag(111) and the observation that bands of
the pure substrate are absent in the experimental window.  The linear
band in silicene/Ag(111) is found to originate primarily from the
substrate and not from Dirac states in the silicene.  To reconcile the
experimental observations  and the calculations for both Ag(111) and
silicene/Ag(111), it is essential that the calculations account for the
$k_\bot$ (photon) dependence of the states.
Our theoretical results provide a consistent explanation of the available
experimental data, and thus resolve the controversy concerning the
(non-)existence of Dirac states in silicene/Ag(111). 

\begin{acknowledgments}   
This work was supported by the U.S. Department of Energy, Office of
Basic Energy Sciences, Division of Materials and Engineering under Award
DE-FG02-07ER46228.
\end {acknowledgments}

\bibliography{references}

\begin{thebibliography}{10}
\bibitem{primitive_Ag111}%
  \BibitemOpen
  \bibinfo {note} {For the (1$\times$1) Ag(111) calculations, a 40-layer slab,
  20 \AA{} vacuum regions, and an 18$\times$18 Monkhorst-Pack mesh were used.}%
  \bibAnnoteFile{Stop}{primitive_Ag111}%
\end{thebibliography}
\bibliographystyle{apsrev4-1}
\end{document}